\documentclass[12pt]{iopart}
\usepackage[english]{babel}
\usepackage{epsfig}

\usepackage{harvard}
\renewcommand{\abstractname}{\textsc{ABSTRACT}}
\makeatletter
\renewcommand{\@seccntformat}[1]{{\csname the#1\endcsname}.\hspace{0.5em}}

\long\def\@makefigcaption#1#2{%
  \vskip\abovecaptionskip
  \sbox\@tempboxa{\textbf{#1.} #2}%
  \ifdim \wd\@tempboxa >\hsize
  \textbf{#1.} #2\par
  \else {
    \global \@minipagefalse
    \hb@xt@\hsize{\hfil\box\@tempboxa\hfil}}%
  \fi
  \vskip\belowcaptionskip}

\renewcommand{\figure}{\let\@makecaption\@makefigcaption\@float{figure}}

\long\def\@maketblcaption#1#2{%
  \vskip\abovecaptionskip
  \begin{center}\small\bf#1\\\normalsize#2\end{center}
  \vskip\belowcaptionskip}

\renewcommand{\table}{\let\@makecaption\@maketblcaption\@float{table}}
\makeatother

\usepackage{endnote} \let\footnote=\endnote
\usepackage{setspace}

\begin{document}

\title[Short-term market reaction]{Short-term market reaction after extreme price changes of liquid stocks}

\author{\'Ad\'am G. Zawadowski$^{1,2}$, Gy\"orgy Andor$^{1}$, and J\'anos Kert\'esz $^{3,4}$}

\address{$^1$Department of Industrial Management and Business Economics, Budapest University of Technology and Economics,
M\H uegyetem rkp. 9, H-1111, Budapest, Hungary}
\address{$^2$Department of Economics, Central European University,
Okt\'ober 6. u. 12., H-1051, Budapest, Hungary}
\address{$^3$Department of Theoretical Physics, Budapest University of Technology and Economics,
Budafoki \'ut 8, H-1111, Budapest, Hungary}
\address{$^4$Laboratory of Computational Engineering, Helsinki University of Technology,
P.O.Box 9400, FIN-02015 HUT, Finland}

{E-mail: zadam@born.phy.bme.hu, andor@imvt.bme.hu, and
kertesz@neumann.phy.bme.hu}

\begin{abstract}
In  our empirical study, we examine the price of liquid stocks
after experiencing a large intraday price change using data from
the NYSE and the NASDAQ. We find significant reversal for both
intraday price decreases and increases. The results are stable
against varying parameters.  While on the NYSE the large widening
of the bid-ask spread eliminates most of the profits that can be
achieved by a contrarian strategy, on the NASDAQ the bid-ask
spread stays almost constant yielding significant short-term
abnormal profits. Furthermore, volatility, volume, and in case of
the NYSE the bid-ask spread, which increase sharply at the event,
decay according to a power-law and stay significantly high over
days afterwards.
\end{abstract}

\maketitle\thispagestyle{empty}
\clearpage

\pagestyle{plain}  \addtocounter{page}{-1}

This paper focuses on intraday market reaction to stock price
shocks. It is only recently that such minute-to-minute analysis of
stock prices has been made possible by the fast improvement of
computers.\footnote{Intraday market reaction for index futures
following large price changes at the opening of the market is
analyzed by \citeasnoun{fung}. However, the small amount of data
existing for index futures only allows the investigation of small
price changes, thus the conclusions are not robust.} Detailed
studies have been devoted to intraday reaction on interest rates
and foreign exchange markets following macroeconomic announcements
(e.g. that of \citeasnoun{ederington}). A thorough empirical study
of the intraday reaction to price shocks on stock markets is still
missing.

Many studies in the past have been devoted to investigating the
abnormal returns following large price changes. In the vast
majority of cases research is focused on daily price decreases of
at least 10\%. Significant price reversal is found during the
first 3 post-event days on all stock markets investigated:
overreaction is found on the NYSE (New York Stock Exchange) and
the AMEX (American Stock Exchange) by
\citeasnoun{bremer91}.\footnote{ Overreaction is found on other
markets as well: \citeasnoun{cox} find significant overreaction on
the TSE (Tokyo Stock Exchange), and \citeasnoun{robertson} in
Johannesburg.} On the other hand, in case of large price increases
either no overreaction is found (\citeasnoun{bremer97} on the TSE)
or the size of the price reversal is much less robust than those
following price decreases (\citeasnoun{atkins} on the NYSE).

In contrast to long-term overreaction (with a time span of years),
such as that found by \citeasnoun{debondt}, which may be at least
partially attributed to a mismeasurement of the risk as described
by \citeasnoun{fama98}, short term pricing errors are much more
robust. In case such short-term overreaction yields abnormal
profits it poses a major threat to the hypothesis of market
efficiency. It is thus crucial to investigate the transaction
costs a trader implementing a contrarian trading strategy faces.
The most important source of transaction costs is the bid-ask
spread: both \citeasnoun{cox}, and \citeasnoun{park} compare the
size of overreaction with that of the bid-ask spread after large
daily price changes, and conclude that no significant profits can
be achieved by arbitrageurs since the anticipated profit of a
contrarian trading strategy does not significantly exceed the
transaction costs: i.e. the reversal for daily data is significant
statistically but not economically.

Thus, the pure existence of an overreaction does not pose a threat
to market efficiency, although a full explanation of the
phenomenon is still to be found. In spite of the fact that no
abnormal profits can be made by external traders (those not owning
stocks suffering serious one day losses) it still remains a puzzle
why investors sell after large price decreases when a significant
rebound is to be expected in the following days.

\citeasnoun{cox} show that on the NYSE the size of the
overreaction after large daily price drops had diminished to zero
until 1991. While in 1963--67 a significant rebound of 1.87\% was
to be expected on the first three days following the event day,
the reversal sank to 0.06\% in 1987--1991, which is not
significant any more. This finding may as well imply that
overreaction is a sign of market inefficiency which was corrected
by market participants after being revealed.

The studies mentioned above all deal with daily close-to-close
price changes. These price changes are due at least partially to
news received by the traders during this 24 hour period, although
\citeasnoun{cutler} show, that in many cases it is hard to trace
back large changes in the S\&P 500 index to one particular event.
Such a time interval may be too long, since many events can take
place during one day. Some events may take place within a day and
thus cannot be unveiled studying daily data. Recent research of
intraday data by \citeasnoun{busse} shows that new information is
incorporated in stock prices within 5--15 minutes. This is in full
correspondence with the fact that autocorrelation of stock prices
diminishes to zero in about the same time (\citeasnoun{wood}).
Hence we might as well expect that it is worthwhile to examine big
intraday price changes of the length from a couple of minutes to a
couple of hours. For example \citeasnoun{fair} introduces a method
in order to find large intraday changes in the S\&P 500 index
which occur within 5 minutes. Instead of an index we concentrate
on large price changes in the price of individual stocks, and
follow a similar, but somewhat more thorough approach when
searching for intraday price shocks, focusing on large price
changes and potential reversal within the trading day. Searching
for the cause of the extreme price changes, although interesting,
is not the task of our study and is not relevant from the point of
testing weak-form market efficiency.

Intraday price changes are interesting not only because of the
possibility of arbitrage but for pure academic reasons too: the
whole price discovery process takes place within the active
trading period. The understanding of this process is not possible
without following the exact intraday transaction bid and ask price
evolution. \citeasnoun{schreiber} point out the importance of such
minute-to-minute empirical investigation which could not yet be
carried out when their paper was written in 1986: now we have the
data and computers have the capability of doing such intraday
analysis.

Recently much research has been devoted to the price discovery
process using high resolution data.\footnote{See e.g.
\citeasnoun{hres1}; \citeasnoun{hres2}; \citeasnoun{hres3}.} The
dynamics of the whole limit order book has been investigated
thoroughly during the past years.\footnote{See e.g.
\citeasnoun{lord1}; \citeasnoun{lord2}; \citeasnoun{lord3};
\citeasnoun{lord4}.} Here we focus on extremal events in the hope
that the dynamics after them will contribute to revealing the
complex mechanism of price formation.

In case of examining intraday price changes we use minute price
data, thus we only examine liquid stocks: those which are traded
minute-to-minute during the trading day. This restriction is
useful in the sense as well that liquid stocks are less exposed to
bid-ask effects (infrequent trading), and thus the transaction
price gives us more direct information on the price traders assume
to be appropriate for the given stock. Additional information on
the minute-to-minute trading around the event can be obtained by
studying the minute volatility and the trading volume which are
also subjects to our studies. We consider two different markets in
order to point out possible differences due to the various trading
mechanisms. Indeed, as we show later, there are significant
differences in the behavior of the NYSE and the NASDAQ.

The paper is organized as follows. We describe the dataset and the
methodology of finding events in Section \ref{data}. Our empirical
findings are presented and discussed in Section \ref{find}.
Section \ref{conc} contains our conclusions.

\newcommand{\runinheader}[1]{\noindent\textbf{#1}:\rule{0mm}{8mm}\hspace{2mm}}

\section{ Data and Methodology \label{data}}

The primary dataset used is the TAQ (Trades and Quotes) database
of the NYSE for the years 2000-2002. The TAQ database is that
supplied by the NYSE: it includes all transactions and the best
bid and ask price for all stocks traded on the NYSE. In our
sample, we include all stocks which were traded anytime during the
observed period. After adjusting for dividend payments and stock
splits, a minute-to-minute dataset is generated using the last
transaction, and the last bid and ask price during every minute.
If no transaction takes place for a minute (which is rare for
liquid stocks on the NYSE), the last transaction price is regarded
as the price. Determining the beginning and the end of the trading
day is done using the dataset: the first trading minute for a
given stock on a given day is the minute during which it was first
traded that day, the last minute is the last minute for which the
DJIA was calculated but not later than 16:00. \footnote{The first
trading minute according to this methodology is usually between
9:30 and 9:35. Some trading occurs after 16:00 but it is omitted
for convenience.} We define liquid stocks as those for which at
least one transaction was filed for at least 90\% of the trading
minutes of the stocks included in the DJIA during the 60 pre-event
trading days.

The secondary dataset is the TAQ database of the NASDAQ. We
include all NASDAQ stocks traded on the first trading day of 2000.
Unfortunately no dividend and stock splits information is
available in the TAQ database for the NASDAQ stocks. However,
intraday analysis is not affected by stock splits. In case of the
NASDAQ determining the opening minute is straightforward, since
the trading is computer based.\footnote{The opening second of the
NASDAQ is 9:30:00 and the closing is 16:00:00 ET.} A problem
arises when using NASDAQ data: there are often singular
transactions filed at a price outside the bid-ask spread
(sometimes even 4-8\% from the mean bid-ask price). Since these do
not represent a change in investor sentiment they are to be
excluded from the sample. Thus the specified trigger levels have
to be surpassed by not only the intraday transaction price change
alone but by the change in the mean of the bid and ask price as
well.

We do not include events in our sample for such stocks where the
price tick is high compared to nominal price either: thus only
stocks with a nominal price over 10 USD are studied. Since we
study the intraday reaction to large price shocks we first have to
define what "extreme price changes" mean. Here we restrict
ourselves to pure intraday price changes: large changes at the
beginning of the day (close-to-open) do not seem to show any
extraordinary effects which are stable to varying the trigger
parameters.

\subsection{Defining large intraday 15-minute price changes \label{iday}}

We use a combined trigger to find the intraday events. Two trivial
methods are at hand:

 \noindent {\sl 1. Absolute filter}: using this first method we
look for intraday
  price changes bigger than a certain level of 2-6\% price change
  within 10-120 minutes. In this case we have to face several problems. Most of
  the events we find occur during the first or last couple of
  minutes of the trading day because of the U-shape intraday volatility distribution
  of  prices (see \citeasnoun{wood}). These events represent
  the intraday trading pattern rather than extreme events. Another problem is that
  a 4\% price jump e.g. may be an everyday event for a volatile stock while
  an  even smaller price move may indicate a major event in case of
  a low volatility stock.

  \noindent {\sl 2. Relative filter}: in case of this  second method we measure the average intraday
  volatility as a function of trading time during the day: this means measuring the
  U-shape intraday 10-120-minute volatility curve
  (length chosen corresponding to the length of the price drop we are going to study)
  for each stock prior to the event. \footnote{This daytime adjustment of volatility is based on
  the idea of \citeasnoun{liu}}
  We define an event as a price move exceeding 6-10 times the
  normal volatility during that time of the day. The problem in
  case of this method is the following: since price moves are very
  small during the noon hours, the average volatility for the 60
  pre-event days in these hours may be close to zero, i.e. a small
  price movement (a mere shift from the bid price to the ask price
  for example) may be denoted as an event. In this case the events
  cluster around the noon hours and no events are found around the
  beginning and the end of the trading day.

The best solution for localizing events is a combined one. Using
the relative filter and absolute filter together, we can eliminate
the negative effect of both filters and combine their advantages.
Thus an event is taken into account if, and only if, it passes
both the relative and the absolute filter. We adjust the absolute
and relative filter so as to achieve that events are found
approximately evenly distributed within the trading day. In
addition we omit the first 5 minutes of trading because we do not
want opening effects in our average. We omit the last 60 minutes
of trading as well because, as shown in Subsection
\ref{find}.\ref{intr}, the major reaction after the price shock
takes place during the 30-60 minutes after the end of the price
change, and we would like to focus on the intraday price reaction
before the market closes.

In order to be able to observe the exact price evolution after the
intraday event, it is crucial to localize the events as precisely
as possible. Since some price changes may be faster than others we
allow shorter price changes than the given 10-120 minutes as well.
For example if when looking for 60 minute events, the price change
already surpasses the filter level in e.g. 38 minutes then we
assume the price change has taken place in 38 minutes. The end of
the time-window in which the event takes place is regarded as the
end of the price change, and this is the point to which the
beginning of the post-event time scale is set: thus minute 0 is
exactly the end of the earliest (and the shortest of those) time
window for which the price change passes the filter. This method
is constructed so as to ensure that one can definitely decide by
minute 0 whether an event has taken place in the preceding 10-120
minutes (time length depending on the specification of the filter)
or not.

\subsection{Calculating the average}

Choosing the sample of events included in the average is again a
crucial step. For convenience we only include the first event
during a given day for a given stock in the average. A major event
may affect the whole market, thus many stocks may experience a
price shock at the same time. Including all events in the average
would yield a sample biased to such major events. In order to
avoid this bias no events are taken into account which happen
within $T=60$ minutes of the previous event included in the
average. The importance of not letting the post-event time periods
overlap is discussed in detail in Appendix \ref{bias}.

When studying intraday price changes, 1 minute volatility, 1
minute trading volume, and the bid-ask spread are averaged besides
the cumulative abnormal return, and the bid and ask price.
Volatility, trading volume, and the bid-ask spread are measured in
comparison to the average minute-volatility and minute trading
volume of the individual stock during the same period of the
trading day (intraday volume and volatility distributions are
calculated using the average of the 60 pre-event trading days).
This step is important in order to remove the intraday U-shape
pattern of the above quantities from the average, since their
intraday variation is in the same order of magnitude as the effect
itself we are looking for.

When studying the intraday effect (within half an hour to one
hour) during the trading day,  the specific way of calculating
abnormal returns seems unimportant since normal profits up to one
hour can probably be neglected. It may however be the case that
market returns are significantly different from the average during
the post-event hours, thus the CAPM is used to calculate abnormal
returns, where the DJIA index is used as the market portfolio for
the NYSE and the NASDAQ Composite Index  for the
NASDAQ.\footnote{Intraday index data is that of Disk Trading Ltd.}
The $\beta$ values for the individual stocks are calculated using
daily (close-to-close) stock prices and index changes of the 60
pre-event trading days, since minute price changes are so noisy
that no reliable $\beta$ values can be calculated from them. Since
$\beta$ values may change after large price shocks, the post-event
abnormal return is calculated using the $\beta$ computed using the
60 post-event trading days (if there are not enough post-event
days in the data, the pre-event value is used).

Our results can be regarded as significant in case both abnormal
and the unadjusted post-event returns are significant as well. We
can conclude from the results in Section \ref{find} that raw and
abnormal (risk-adjusted) returns are not significantly different,
which is not surprising in the light of the fact described above
that expected returns for such short intervals as 30-60 minutes is
practically zero. Thus we will stay with raw returns when
analyzing the stability of our filter.

T-values are calculated as deviation of mean from null-hypothesis
(0\% in case of returns) over the standard deviation of the mean
in each and every case.

\section{Empirical results \label{find}}

\subsection{ Intraday price reaction \label{intr} }

\begin{figure}
  \caption{\bf{Price reversal after intraday price increases and decreases exceeding 4\% and 8 times the pre-event volatility
  on the NYSE and the NASDAQ}\label{rev}} Events are maximum 60-minute long price drops exceeding 4\% and 8
times the average 60-minute volatility of the same time interval
of the 60 pre-event days. Returns are raw returns. An average of
175 increases and 222 decreases on the NYSE and 273 increases and
215 decreases on the  NASDAQ respectively. Minute 0 corresponds to
  the time when the price change exceeds the combined filter level.

  \setstretch{1.2}
  \centerline{\epsfig{file=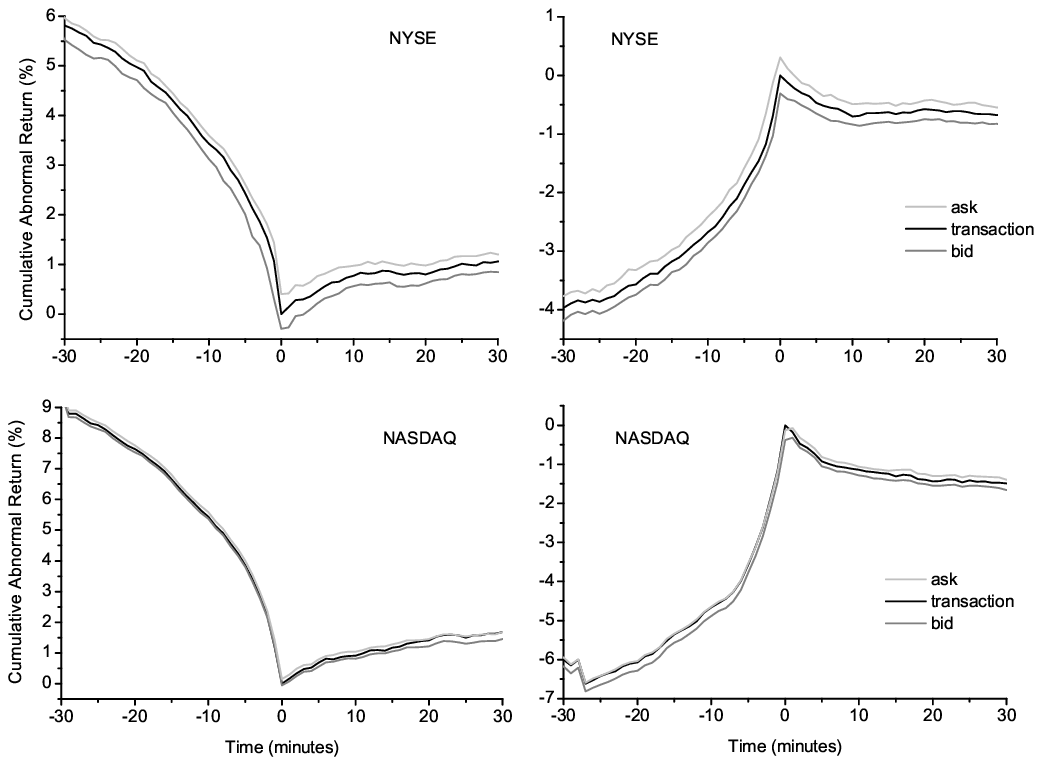,width=17truecm}}
\end{figure}

We present abnormal and raw returns after intraday 60 minute price
shocks in Table~\ref{ti} for the NYSE and Table~\ref{tinasdaq} for
the NASDAQ . The price changes included in the average are in
absolute value all bigger than 4\% and 8 times the average
60-minute volatility of the corresponding 60 minutes during the 60
pre-event trading days. These trigger values were determined using
the NYSE dataset in order to find only extreme events, and to
result in an approximately even distribution of events during the
trading day: Table~\ref{dist} shows the intraday distribution of
events. Events are at least 60 minutes apart, thus only returns
during a time period of maximum 60 minutes should be calculated in
correspondence with  Appendix \ref{bias}. Figure~\ref{rev} shows
the exact evolution of transaction, bid, and ask price around the
event for NYSE and NASDAQ stocks: there is a clear reversal in
case of both decreases and increases. The phenomenon of the
bid-ask bounce, which \citeasnoun{cox} claim is a substantial
cause of daily price reversals on the NYSE, cannot be observed for
intraday price reversal of liquid NYSE stocks, neither for NASDAQ
stocks.

\begin{table}[ht]
  \caption{ Price evolution after intraday price decreases and increases exceeding
 4 \% and 8 times the pre-event volatility  for liquid stocks on the NYSE  \label{ti}}
Events are maximum 60-minute long price drops exceeding 4\% and 8
times the average 60-minute volatility of the same time interval
of the 60 pre-event days. Abnormal returns (AR) are calculated by
subtracting beta times the corresponding index return (DJIA) from
the raw return (RR) of the stock's price on the basis of TAQ
database. Pre- and
  post-event betas are calculated from 60 daily returns before and after the
  event. An average of 175 increases and 222 decreases at least 60 minutes from each other and at least
  60 minutes before market closure. Minute 0 corresponds to
  the time when the price change exceeds the  filter level.
  T-values in parenthesis.
\vskip0.5cm
  \setstretch{1.2}
  \begin{tabular}{ll@{\hspace{15mm}}llll} % Note:  does not use vertical lines
    \hline                                %  does not use double hlines

   from & to &  \multicolumn{2}{c}{increase}& \multicolumn{2}{c}{decrease} \\
     \multicolumn{2}{l}{time (minutes)} & AR  & RR  &AR  & RR    \\
    \hline
    -120&   -60&    -1.825\%**&  -2.147\%**&  -0.978\%**& -1.046 \%**\\
    &   &   (4.60)  & (5.64)  & (2.41)  & (2.51)   \\
      -60  & 0 &  3.085\%** & 3.247\%** & -7.197\%** & -7.56\%**\\
      &      &  (8.17) & (8.33)  & (16.95) & (18.05)\\
     0  &  10   &  -0.725\%** & -0.698\%** & 0.842\%** &  0.777\%**\\
      &     &  (6.72) & (6.46)  & (6.24) & (5.61))\\
     0    &  30   &  -0.728\%** & -0.678\%** & 1.132\%** &  1.061\%**\\
      &      &  (5.28) & (4.58)  & (5.91) & (5.40) \\
     0    &  60   &  -0.629\%** & -0.561\%** & 1.125\%** & 1.022 \%**\\
      &      &  (4.06) & (3.21)  & (4.28) & (3.69) \\
        60  &  120   &  -0.120\% & -0.059\% & 0.013\% &  0.334\%\\
      &      &  (0.70) & (0.31)  & (0.05) & (1.30) \\
    \hline                                  %  does not use double hlines
  \end{tabular}
  \vspace{0.2cm}

* Indicates mean significantly different from zero at the 95\%
level.

** Indicates mean significantly different from zero at the 99\%
level.
  \vspace{1cm}

\end{table}

\begin{table}[ht]
  \caption{ Price evolution after intraday price decreases and increases exceeding
 4 \% and 8 times the pre-event volatility  for liquid stocks on the  NASDAQ  \label{tinasdaq}}
Events are maximum 60-minute long price drops exceeding 4\% and 8
times the average 60-minute volatility of the same time interval
of the 60 pre-event days. Abnormal returns (AR) are calculated by
subtracting beta times the corresponding index return (NASDAQ
Composite) from the raw return (RR) of the stock's price on the
basis of TAQ database. Pre- and
  post-event betas are calculated from 60 daily returns before and after the
  event. An average of  273 increases and 215 decreases at least 60 minutes from each other and at least
  60 minutes before market closure. Minute 0 corresponds to
  the time when the price change exceeds the  filter level.
  T-values in parenthesis.
\vskip0.5cm
  \setstretch{1.2}
  \begin{tabular}{ll@{\hspace{15mm}}llll} % Note:  does not use vertical lines
    \hline                                %  does not use double hlines
   from & to & \multicolumn{2}{c}{increase} &\multicolumn{2}{c}{decrease} \\
     \multicolumn{2}{l}{time (minutes)}  &AR  & RR  &AR  & RR  \\
    \hline
    -120&   -60&     -1.076\%**&  -1.566\%**&  -1.377\%**&  -1.587\%**\\
    &   &   (2.41)  & (3.34)  & (2.97)  & (3.48)  \\
      -60  & 0 &   5.344\%** & 6.215\%** &-10.98\%** &  -12.23\%**\\
      &      &   (6.90) & (7.54)  & (10.27) & (11.32)\\
     0  &  10   &  -1.241\%** & -1.140\%** & 1.063\%** &  0.916\%**\\
      &     & (8.69) & (7.31)  & (3.27) & (2.79)\\
     0    &  30   & -1.507\%** & -1.486\%** & 1.832\%** &  1.685\%**\\
      &      &    (8.13) & (7.18)  & (5.47) & (4.73) \\
     0    &  60   &    -1.683\%** & -1.492\%** & 2.149\%** &  1.931\%**\\
      &      &   (7.86) & (6.26)  & (5.91) & (5.25)\\
        60  &  120   &  0.020\% & 0.027\% & 0.170\% &  0.553\%\\
      &      &    (0.12) & (0.14)  & (0.59) & (1.64)\\
    \hline                                  %  does not use double hlines
  \end{tabular}
  \vspace{0.2cm}

* Indicates mean significantly different from zero at the 95\%
level.

** Indicates mean significantly different from zero at the 99\%
level.
  \vspace{1cm}

\end{table}

\begin{table}[ht]
  \caption{Intraday distribution of the end of 60-minute events exceeding 4\% and 8 times the normal volatility\label{dist}}

Time periods correspond to 30 minute periods from 9:30 to 16:00.
No events are taken into account after 15:00 (one hour before
market closure). Events during the first hour are rare since most
events indeed last 60 minutes. Events are approximately evenly
distributed within the trading day for both NASDAQ and the NYSE.
 \vskip0.5cm
  \setstretch{1.2}
  \begin{tabular}{l@{\hspace{2mm}}lllllllllll} % Note:  does not use vertical lines
    \hline                                %  does not use double hlines
    period & 9:30 &10:00 &10:30 &11:00 &11:30 &12:00& 12:30&13:00&13:30&14:00& 14:30\\
    \hline
    NYSE &&&&&&&&&&&\\
    decreases &2&21&25&22&17&33&30&15&16&21&20\\
    increases  &1&10&16&15&24&14&25&18&22&13&17\\
    NASDAQ &&&&&&&&&&&\\
    decreases  &4&15&20&23&35&22&29&19&16&19&13\\
    increases  &4&12&17&43&26&23&32&38&29&29&20\\
    \hline                                  %  does not use double hlines
  \end{tabular}
  \vspace{1cm}

\end{table}

Investigating the post-event returns we see a significant reversal
for transaction prices in 10-60 minutes after the event. The
reversal seems faster after price increases, while for price
decreases it is slower. We can infer from Table~\ref{ti} and
Table~\ref{tinasdaq} that the price reversal (although of almost
the same size after both increases and decreases during the first
10 minutes) is bigger after decreases for a 30-60 minutes
interval. The size of the reversal is large if we take into
account that it happens within 30-60 minutes for stocks with high
liquidity.

Abnormal and raw returns are very close to each other on both
markets, which is not surprising once we know that market return
on average is not high within one hour. On the other hand the fact
that abnormal returns are higher than raw returns is somewhat
surprising. Using Table~\ref{ti} and Table~\ref{tinasdaq} we may
conclude that adjusting for market risk does not change our
results, thus we will only calculate raw returns in the remaining
part of the paper since they are more straightforward to
interpret.

\subsection{ Stability of the price reversal \label{estab} }

Although we have shown that there is a significant price reversal
following extreme price changes, it is not yet clear whether the
phenomenon is stable to varying the parameters. In
Table~\ref{kulon} we show the 60 minute post-event returns as a
function of the parameters of both the absolute and the relative
filter. Table~\ref{idok} shows the size of the rebound as the
function of the time during which the large price change took
place.

\begin{table}[ht]
  \caption{60 minute raw returns after intraday 60-minute price drops passing the combined filter for
  liquid NYSE and NASDAQ stocks \label{kulon}}
Events are maximum 60-minute long price drops exceeding 2-6\% and
6-10 times the average 60-minute volatility during the same period
of the day in the 60 pre-event days.
   \vskip0.5cm
  \setstretch{1.2}
  \begin{tabular}{l@{\hspace{15mm}}lll@{\hspace{5mm}}lll} % Note:  does not use vertical lines
    \hline                                %  does not use double hlines
      & \multicolumn{3}{c}{NYSE}&\multicolumn{3}{c}{NASDAQ} \\
       & 6 & 8 & 10  & 6 & 8 & 10 \\
      \% & \multicolumn{3}{c}{times normal volatility} & \multicolumn{3}{c}{times normal volatility} \\

    \hline
    2\% &  0.507\%** & 0.813\%**& 1.077\%** & 1.067\%** &1.932\%**  &3.120\%** \\
    {\sl t-stat}  & (4.60) & (3.69) & (2.59) & (6.75) & (5.44) & (5.08)\\
    {\sl number}   & 628& 277 & 141 & 609 & 223 & 104\\
   \hline
   4\%  &  0.573\%** & 1.022\%**& 1.229\%** & 1.126\%** &1.931\%**  &3.104\%** \\
    {\sl t-stat} & (3.42) & (3.69) & (2.62) & (6.56) & (5.25) & (5.00)\\
    {\sl number}    & 405 & 222 & 126 & 567 &215  &103 \\
   \hline
   6\%  &  0.911\%** & 1.256\%**& 1.317\%** & 1.366\%** &2.063\%**  &3.160\%** \\
    {\sl t-stat}  & (3.15) & (3.04) & (2.11) & (6.22) & (5.06) & (4.76)\\
    {\sl number}    &228 & 144 & 94  & 429 & 193 &96 \\
   \hline                                 %  does not use double hlines
  \end{tabular}
  \vspace{0.2cm}

* Indicates mean significantly different from zero at the 95\%
level.

** Indicates mean significantly different from zero at the 99\%
level.
  \vspace{1cm}
\end{table}

\begin{table}[ht]
  \caption{60 minute raw returns after intraday 10-120-minute price changes passing the combined filter for
  liquid NYSE and NASDAQ stocks \label{idok}}
Events are maximum 10-120-minute long price drops exceeding 4\%
and 8 times the average 10-120-minute volatility during the same
period of the day in the 60 pre-event days.
   \vskip0.5cm
  \setstretch{1.2}
  \begin{tabular}{l@{\hspace{15mm}}llllll} % Note:  does not use vertical lines
    \hline                                %  does not use double hlines

       & 10 & 20 & 40  & 60 & 90 & 120 \\
    \hline
    \multicolumn{7}{l}{NASDAQ up}\\
     0-10 & -0.250\%* & -0.641\%**& -0.874\%** & -1.140\%** &-1.111\%**  &-1.028\%** \\
    {\sl t-stat}  & (1.98) & (5.60) & (6.34) & (7.31) & (5.42) & (4.77)\\
    0-60 &  -0.130\% & -0.617\%**& -0.963\%** & -1.49\%** &-1.835\%**  &-2.067\%** \\
    {\sl t-stat}  & (0.65) & (3.54) & (4.04) & (6.26) & (6.30) & (7.17)\\
    {\sl number}  & 470& 475 & 339 & 273 & 199 &159 \\
   \hline
   \multicolumn{7}{l}{NYSE up}\\
    0-10 &  -0.543\%** & -0.503\%**& -0.609\%** & -0.698\%** &-0.648\%**  &-0.597\%** \\
    {\sl t-stat}  & (3.39) & (3.94) & (5.67) & (6.46) & (5.65) & (5.89)\\
    0-60  &  -0.236\% & -0.043\%& -0.353\%* & -0.561\%** &-1.003\%**  &-1.058\%** \\
    {\sl t-stat} & (0.82) & (0.19) & (1.92) & (3.21) & (4.82) & (5.41)\\
    {\sl number}    &149 & 197 & 209 & 175 & 152 & 131 \\
    \hline
    \multicolumn{7}{l}{NASDAQ down}\\
     0-10 &  0.352\%* & 0.298\%& 0.705\%** & 0.663\%* &1.235\%**  &1.554\%** \\
    {\sl t-stat}  & (1.83) & (1.41) & (2.63) & (1.98) & (2.93) & (3.00)\\
     0-60 &  0.617\%** & 0.773\%**& 1.415\%** & 1.93\%** &2.322\%**  &3.150\%** \\
    {\sl t-stat}  & (2.49) & (3.03) & (4.54) & (5.25) & (5.25) & (5.37)\\
    {\sl number}   &394 & 369  & 283 & 215 & 161 & 127\\
   \hline
    \multicolumn{7}{l}{NYSE down}\\
     0-10 &  0.286\% & 0.600\%**& 0.484\%** & 0.777\%** &0.683\%**  & 0.846\%** \\
    {\sl t-stat}  & (1.27) & (3.51) & (3.34) & (5.61) & (4.72) & (5.05)\\
    0-60&  0.677\% & 0.698\%*& 0.713\%** & 1.022\%** &1.297\%**  &1.322\%** \\
    {\sl t-stat} & (1.59) & (2.09) & (2.71) & (3.69) & (3.76) & (3.53)\\
    {\sl number}    & 155& 212 & 246 & 222 & 176 & 156 \\
   \hline                                 %  does not use double hlines
  \end{tabular}
  \vspace{0.2cm}

* Indicates mean significantly different from zero at the 95\%
level.

** Indicates mean significantly different from zero at the 99\%
level.
  \vspace{1cm}
\end{table}

Our results imply that the phenomenon of overreaction is stable
even when varying parameters. We can infer from Table~\ref{kulon}
that the exact size of the filter does not affect the existence of
overreaction, even though using more stringent filters -- i.e.
only selecting very rare and extreme events -- the rebound gets
larger and larger. Table~\ref{idok} shows us that the length of
the price drop is a crucial parameter, if it is chosen to be too
short results tend to be insignificant. This may be due to the
relative filter: for longer time horizons outliers are less and
less frequent(representing more extreme events) because the
distribution of price changes converges to normal as shown by
\citeasnoun{stanley}. This hypothesis is supported by the sharply
decreasing number of extreme events passing the filter as we
lengthen the time interval of the price drop.

\begin{figure}
  \caption{\bf{The size of 60-minute reversals following large price drops as a function
  of the original 60-minute price drop}\label{lin}}

Data points are averages for different filter levels, see
Table~\ref{ti} for NYSE and Table~\ref{tinasdaq} for the NASDAQ.
\vskip0.5cm
  \setstretch{1.2}
  \centerline{\epsfig{file=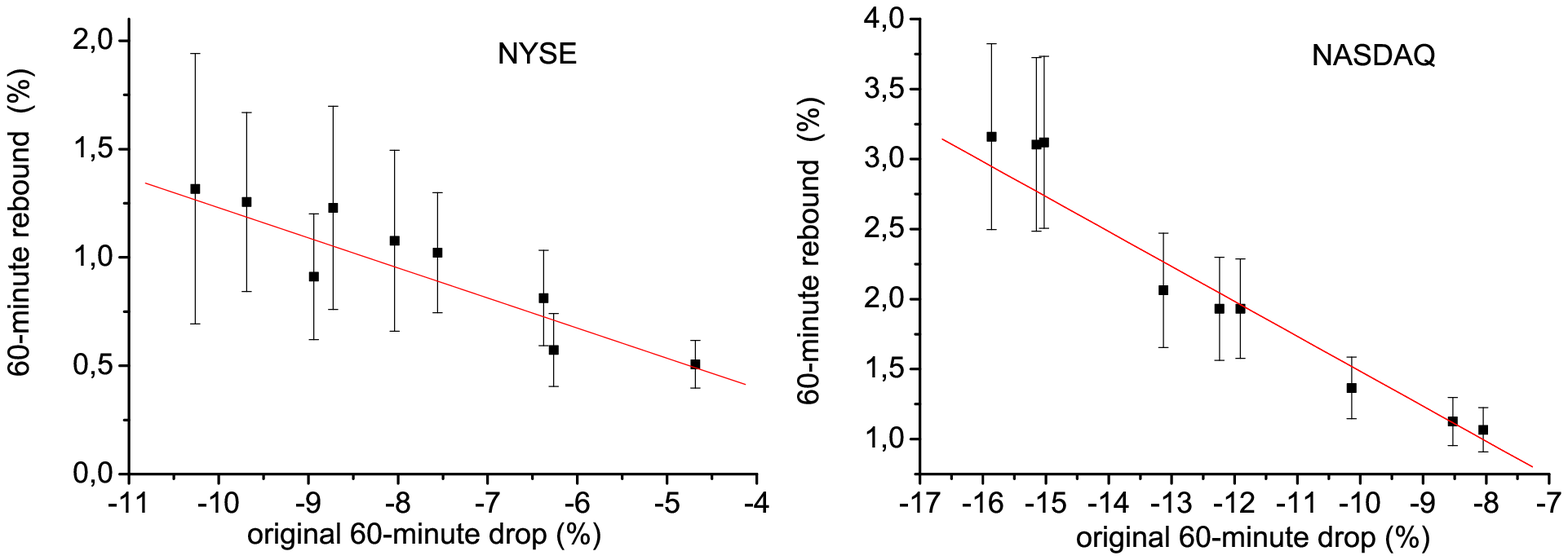,width=17truecm}}
\end{figure}

The above results are indeed puzzling, we may as well call it the
"intraday reversal puzzle". Although we restrict ourselves to the
demonstration of empirical facts in this paper we give two
possible explanations. One possibility is to see this puzzle as a
clear proof of behavioral finance in the short run: chartist
traders who deal irrationally simply overreact the actions of
fundamentalist traders and a pricing error arises which is then
later reversed. This phenomenon is shown to exist in simulation
models of the stock market by \citeasnoun{sajat}. Another
theoretical approach to the puzzle is that reversals are a natural
and rational consequence of the coexistence of informed and
uninformed traders on the market: \citeasnoun{gabaix} show that if
it is unsure whether a large price move was caused by informed or
by uninformed traders this rationally leads to a price reversal.
Both studies imply that there should be a linear relationship
between the size of the original price drop and the size of the
rebound. We can check this hypothesis using the results from
Table~\ref{kulon}. Figure~\ref{lin} indeed shows a positive
relationship between the average size of the 60-minute rebound and
the average size of the original 60-minute price drop on both
markets. The slope of the fitted regression line (weighted by
measurement errors) is $-0.139\pm0.046$ for the NYSE and
$-0.249\pm0.042$ for the NASDAQ (both results significant at the
99\% level). The slope of the fitted line should although be
handled with care since the data points represent averages which
are not independent: more of the averages may contain the same
large event. Observing the fitted line on Figure~\ref{lin} one can
conclude that the linear relationship is in agreement with the
data although a tendency to larger than linear reaction for large
drops, especially for the NASDAQ data, cannot be excluded.
Nevertheless the claim that the larger the initial price change,
the larger the reversal seems generally justified. One may rightly
ask why we restrict ourselves only to investigate this
relationship using the averages, why not regress the rebound on
the original drops for individual. The answer is that it is
impossible to measure exactly how large the original price changes
are, where they start and where they end. We must therefore
restrict ourselves to using different filter levels to be able to
find price changes of different size.

\subsection{ Evolution of the volatility and volume   \label{evol} }

\begin{figure}
 \caption{\bf{Evolution of minute volatility, minute dollar volume, and the bid-ask spread on the NYSE and the NASDAQ after intraday price
  decreases exceeding 4\% and 8 times the pre-event volatility }  \label{volvol}}
  Events are maximum 60-minute long price drops exceeding 4\% and 8
times the average 60-minute volatility of the same time interval
of the 60 pre-event days. Minute volatility, minute volume and the
bid-ask spread are given as a percentage of the 60 pre-event day
average of the same minute during the day. An average of 222
events on the NYSE and 215 on the NASDAQ. All events in one sample
are at least $T=60$ minutes from each other.

  \setstretch{1.2}
  \centerline{\epsfig{file=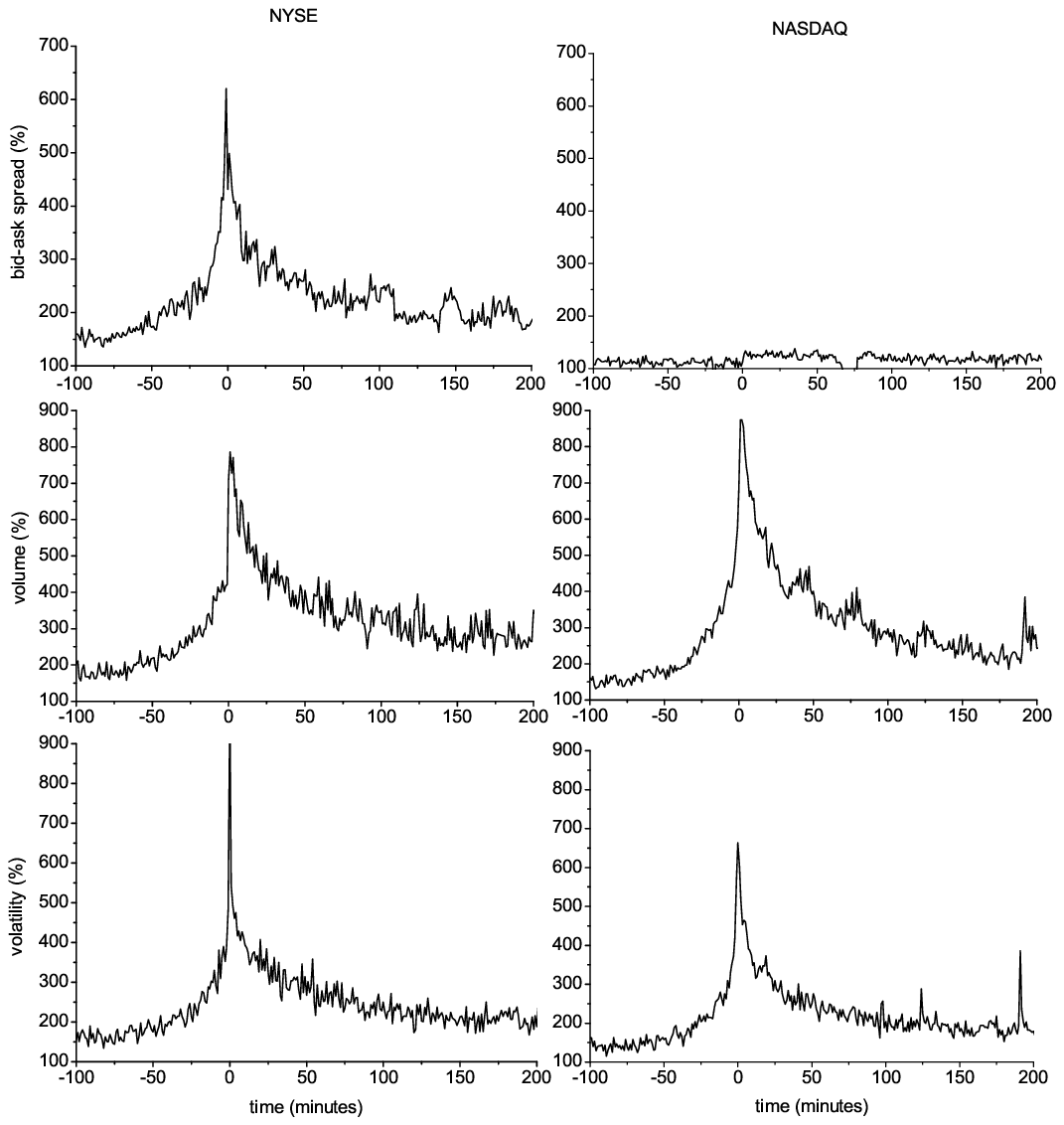,width=15truecm,height=17truecm}}
\end{figure}

Before turning to the question whether these predictable price
patterns are exploitable, let us examine the evolution of
volatility, volume, and the bid-ask spread after the event.
Figure~\ref{volvol} shows only the evolution after price decreases
on the NYSE and the NASDAQ, which is practically the same for
price increases. The volatility is the trigger event itself, thus
it increases sharply during the 60 minutes of the price drop. The
transaction dollar volume increases sharply at the event as well
on both markets, up to 8-9 times of its value during the pre-event
days. Both the volume and volatility (and in case of the NYSE the
bid-ask spread) decrease only very slowly after the event.

\begin{figure}
 \caption{Decay of volatility and volume on the NASDAQ\label{hfv}}
Post-event evolution of excess volatility and excess volume on
log-log scale for NASDAQ stocks experiencing a 10 minute long
price drop exceeding 8 times the pre-event daytime adjusted
volatility and 4\%. While volatility shows power-law decay, volume
does not.
 \vskip0.5cm
  \setstretch{1.2}
  \centerline{\epsfig{file=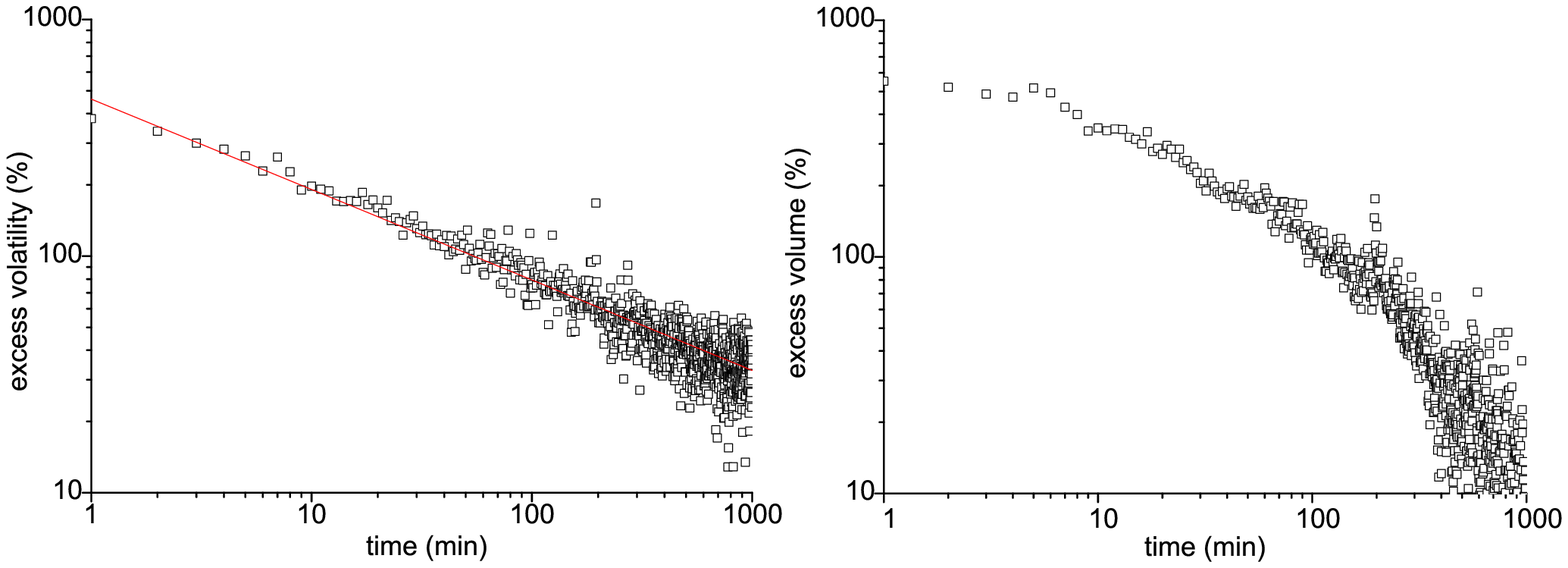,width=17truecm}}
\end{figure}

\begin{table}[ht]
  \caption{Exponent of fitted power-law decay for volatility\label{hfvt}}
Post-event evolution of excess volatility during the first 1000
minutes  can be well fitted by a power-law. The reported exponents
are those obtained by simple OLS on minute data for different
filter parameters on both markets for both increases and
decreases.
   \vskip0.5cm
  \setstretch{1.2}
  \begin{tabular}{l@{\hspace{15mm}}lllll} % Note:  does not use vertical lines
    \hline                                %  does not use double hlines
    relative filter& 8&6&10&8&8\\
    absolute filter&4&2&6&4&4\\
    event length   & 60 & 60 & 60  & 10 &120  \\
    \hline
    NYSE up    &-0.27&-0.33&-0.40&-0.31&-0.25\\
    NYSE down  &-0.35&-0.25&-0.38&-0.31&-0.34\\
    NASDAQ up  &-0.37&-0.35&-0.41&-0.39&-0.37\\
    NASDAQ down&-0.32&-0.35&-0.32&-0.38&-0.40\\
     \end{tabular}
  \vspace{0.2cm}
  Mean errors are all below 0.01
\end{table}

\begin{table}

  \caption{Power exponent of the decay of autocorrelation of volatility\label{cegek}}
   Measured for some of the most liquid NYSE stocks (Jan. 2000--Sep. 2002)
  between 1--1000 minutes \label{power}
  \begin{tabular}{lll}
    \hline
    { symbol} & { name} & { power exponent} \\

    \hline
    GE  & General Electric & $-0.180 \pm 0.015$  \\
    AOL  & AOL Time Warner &   $-0.171 \pm 0.014$ \\
    C  & Citigroup Inc. &  $-0.138 \pm 0.016$  \\
    HD  & Home Depot &   $-0.160 \pm 0.019$ \\
    T  & AT\&T Corp. &   $-0.183 \pm 0.019 $ \\
    IBM  & Int'l Business Machines &   $-0.157 \pm 0.024$ \\
    NOK  & Nokia &   $-0.202 \pm 0.015$ \\
    \hline
  \end{tabular}
  \vspace{0.5cm}
\end{table}

The decrease of post-event excess volatility can in fact be well
fitted by a power-law as shown in Figure \ref{hfv}. On the other
hand it ce=an be inferred from the figure that the decay of excess
volume cannot be described by a power-law. Graphical analysis
reveals that none of the variables decay exponentially. We measure
the exponent of the power-law for the both markets for a variety
of filter levels. From Table \ref{hfvt} we can conclude that the
exponent is between 0.25-0.35 on the NYSE and 0.35-0.40 on the
NASDAQ. In the vast majority of the cases the power-law gives a
very good fit. As we have already done for a small number of
filter levels (\citeasnoun{sajatexo}) we can  compare the exponent
of decay with that of the decay of the autocorrelation of
volatility. \citeasnoun{liu} show that the autocorrelation of
volatility decays according to a power-law with an exponent of
-0.3, pointing out at the same time that the exponent is lower for
the first 1000 minutes. Since the authors do not report results
for the decay of autocorrelation during the first 1000 minutes
(only results from detrended fluctuation analysis) we compute the
exponent of the power-law decay for some selected stocks using the
same method as \citeasnoun{liu} on data from 2000-2002. The
results are presented in Table \ref{cegek} showing that extreme
events decay faster than the autocorrelation. A possible
explanation is that large shocks are more likely to be exogenous
than all fluctuations incorporated in the ACF
(\citeasnoun{sajatexo}).

In case of the bid-ask spread on the other hand, which is a major
source of transaction costs, we find that it stays virtually
unchanged on the NASDAQ  but increases to six times its pre-event
value on the NYSE (decreasing according to a power-law too). This
fact, although in full agreement with the finding of
\citeasnoun{chan} that the bid-ask spread does not vary
substantially on the NASDAQ during the trading day, implies that a
contrarian strategy following the extreme price change has much
lower costs on the NASDAQ than on the NYSE.

\subsection{Profitability of contrarian strategies}

To test weak-form market efficiency we have to check whether the
large and highly significant reversals on both markets are
exploitable, i.e. whether a contrarian strategy yields abnormal
profits. Indeed if we calculate the abnormal profit a contrarian
trading strategy may achieve during the post-event 30-60 minutes
by buying at the ask price at minute 0 and selling at the bid half
to one hour later it is significant on both markets
(Table~\ref{win}).

\begin{table}[ht]
  \caption{Intraday profitability of contrarian strategies\label{win}}
Profitability of buy and hold for 60 minutes strategies. Buying
price is the best ask price, selling price the best bid price.
 \vskip0.5cm
  \setstretch{1.2}
  \begin{tabular}{l@{\hspace{5mm}}llll} % Note:  does not use vertical lines
    \hline                                %  does not use double hlines
    minutes & NASDAQ 60-min & NASDAQ 120-min & NYSE 60-min &NYSE 120-min\\
    \hline
     0-30 &1.300\%** &2.093\%** &0.441\%* &0.744\%**  \\
    t-stat  & (3.24)& (3.11)& (2.18)& (2.85)\\
    0-60 &1.599\%** &2.686\%** &0.439\% &0.919\%**  \\
    t-stat  & (3.87)& (4.02)& (1.53)& (2.50)\\
    2-60 &1.268\%** &2.149\%** & 0.257\% &0.695\%*  \\
    t-stat  & (3.84)& (4.17)& (0.92)& (1.94)\\

    \hline                                  %  does not use double hlines
  \end{tabular}
  \vspace{1cm}

* Indicates mean significantly different from zero at the 95\%
level.

** Indicates mean significantly different from zero at the 99\%
level.
\end{table}

Omitting all other transaction costs, except for the bid-ask
spread, the a trader following a contrarian strategy  achieves
significant profits on the NASDAQ: after 120-minute price drops
2.686\% (4.02) as average of 159 events during the time period
2000-2002. The profits are lower -- or in some cases even
insignificant -- in case of the NYSE because of the widened
bid-ask spread. One may argue that these profits can only be
acquired by traders who are amazingly fast, but that is not the
case. If the trader implementing the contrarian strategy is
relatively slow and is only able to buy the stocks two minutes
after the event he still receives a profit of 2.149\% (4.17)
between minutes 2 and 60 on the NASDAQ. Since no events included
in the sample happen after one hour before market closure these
profits are realized in one hour within the trading day, thus they
are unlikely to be seriously affected by mismeasurement of risk.
Other trading costs paid by brokers on the NASDAQ are probably not
high enough to eliminate these abnormal profits. One must however
state that the amount of gain may be limited because limited
number of shares are available at the best bid and the best ask
price, but profits are significantly larger than 0 (even if not of
great absolute value). Further research using the order book at
the whole depth and looking at the contracts offered by liquidity
providers to their customers may reveal the exact profit which can
be achieved by the above trading strategy.

The number of stocks which pass the liquidity filter, i.e. the
number of stocks which have to be tracked each day during the
trading, is 101-144 for the NASDAQ and 47-252 for the NYSE.
Sophisticated trading software can in theory monitor this number
of stocks throughout the day signaling when an event has taken
place, when to buy and when to sell. Further studies should
examine the exact profitability, if any, of such strategies taking
into account all other costs of transaction and the costs of
monitoring the prices of liquid stocks.

\section{ Conclusions \label{conc}}

In  our empirical study we examine the price of liquid stocks
after experiencing a large 60-minute price change using data from
the NYSE and the NASDAQ. Focusing on short-term reaction  we find
significant overreaction on both markets in the subsequent 30-60
minutes. Stability analysis confirms our findings. This implies
there is a short-term market inefficiency which we name "intraday
reversal puzzle". and one of the main goals of our paper is to
call the attention to this empirical fact. Presently, we can only
speculate about the origin of this phenomenon. It may be due to
behavioral trading during the period of rapidly changing price or
to some kind of interaction between informed and uninformed
traders. We show that the size of the reversal is positively
dependent on the initial price change and compatible with a linear
assumption, at least for not too large changes.

An interesting aspect is the very different behavior of the
bid-ask spread on the NYSE and on the NASDAQ close to a major
price change: an enormous increase is observed on the former,
while practically no significant change accompanies the event on
the latter (cf. Fig. \ref{volvol}). The origin of this has to be
rooted in the different market mechanisms, and we believe that the
explanation is somewhat paradoxical.  On the NYSE where a single
specialist acts, he/she eliminates a large part of the profit a
contrarian strategy may achieve by simply widening the bid-ask
spread. At the same time, the highly competitive dealership market
on the NASDAQ keeps the bid-ask spread almost constant, thus
giving way to a short-term abnormal profit possibility.

In case of extreme price changes we find that intraday contrarian
trading strategies are profitable even after controlling for the
bid-ask spread. Exact profitability of such strategies should
although be studied further by taking into account further sources
of market friction.

Furthermore we can conclude that volatility, volume, and in case
of the NYSE the bid-ask spread, which increase sharply at the
event, stay significantly high long after the price adjustment has
taken place. In fact the post-event decay of excess volatility
(and in case of the NYSE the bid-ask spread) can be well described
by a power-law. Comparing the exponent with that of the
autocorrelation we can conclude that extreme price shocks decay
faster than price fluctuations on average.

\clearpage

\appendix

\section{ Possible biases when calculating
the average \label{bias} }

\setcounter{section}{1}

When averaging the events, we used a minimum distance T between
two events included in the sample. In this subsection we address
the importance of this issue using a GARCH(1,1) model of minute
index data. The equations of the process are:

$$x_{t+1}=x_t+\eta_{t+1}$$

$$\sigma_{t+1}^2=\alpha_o+\alpha_1(x_t-x_{t-1})^2+\beta_1\sigma^2_t$$

$$\eta_{t+1}\in N(0,\sigma_{t+1})$$

\noindent where $\alpha_o$, $\alpha_1$, and $\beta_1$ are the
parameters of the process.  We model the minute price evolution of
the market using the following parameters measured on General
Electric on the NYSE during the years 2000-2002. Parameters are
estimated on the time series of daytime-adjusted returns (daytime
averages are measured on the full three-year period) for 295120
data points. The measured parameters using EVIEWS econometrics
software package are: $\alpha_o=0.0168$, $\alpha_1=0.0663$, and
$\beta_1=0.933$. Thus one time-step of the simulation corresponds
to one minute in case of real-market data.

We investigate relatively small sample (150 events) t-tests for
the significance of returns in this time series model. In
Table~\ref{tbi} we present returns around one minute price drops
of at least 8 (which corresponds to 8 times the usual
minute-volatility during the same period of the day) with a
minimum distance of $T=60$. With this trigger value we get sample
events every 403.7 time steps (minutes) which is in the order of
magnitude of that measured on real data (e.g. 356 price drops
included in the average in 3 years trading time on the NASDAQ). A
GARCH(1,1) process does not exhibit serial correlations (that is
why we choose one-minute price drops in the case of the model),
thus in theory no reversal may take place and no pre-event drops
or increases either. Any significant price patterns measured are
due to sample selection bias.

\begin{table}[ht]
  \caption{The effect of cross sample correlations on significance tests in a model simulation using a
  GARCH(1,1) process \label{tbi}}
Average of $1000*150$ price drops of a GARCH(1,1) process with a
distance of at least $T=60$ between sample events. Returns and
t-values are that of the overall averages. T-average is the
average of t-values, abs(t) average of the absolute t-values of
1000 samples of 150 events each. {\sl PS} is the percentage of
small (150 event) samples where t-value indicates significant
deviation from zero return at the confidence level of 98\%. In
case of uncorrelated sample events this value should be $PS\approx
2\%$.

\vskip0.5cm
  \setstretch{1.2}
  \begin{tabular}{ll@{\hspace{5mm}}lllll} % Note:  does not use vertical lines
    \hline                                %  does not use double hlines
      \multicolumn{2}{c}{time}  & return & (t-value) & t-average & abs(t) average  & PS \\
     from & to &&&&&\\
    \hline
    -181& -121  &-0.186 & (-1.28)& -0.065 & 0.812 &  1.9\% \\
    -121& -61  & -0.439**& (-2.93)& -0.133 & 0.812 &  2.6\% \\
    -61& -1  & 8.72**& (56.20)& 2.05 & 2.07 &  37.8\%$^{++}$\\
    -1&  0   & -10.76**& (-789.6)& -39.5 & 39.5 &  100.0\%$^{++}$ \\
    0& 10   & 0.00942& (0.145)& 0.009 & 0.812 &  2.4\% \\
    0& 60   & -0.0132& (-0.085)& -0.013 & 0.799 &  1.7\% \\
    60& 120  & -0.172&(-1.13) & -0.062 & 0.824 &  2.4\% \\
    0& 390  &-0.0175 & (-0.047)& -0.034 & 1.771 &  29.3\%$^{++}$ \\
    0& 1000  & -0.599 &(-1.094) & -0.035 & 2.492 &  46.3\%$^{++}$ \\
    \hline                                  %  does not use double hlines
  \end{tabular}
  \vspace{0.2cm}

** Indicates large sample mean return significantly different from
0 at the 99\% level.

$^{++}$ Indicates PS significantly different from 2\% at the 99\%
level.

  \vspace{1cm}

\end{table}

It can be inferred from the data that calculated t-statistics for
returns during longer time periods than T are biased (although the
returns themselves are not). Because of the spurious cross-sample
correlation we get a t-value indicating significant deviation (at
the 98\% level using a two-sided test) from the fact that returns
are zero in much more than 2\% of the cases (see last column in
Table~\ref{tbi}). Thus returns on real-market data of time periods
longer than T should not be tested by the t-statistics. This holds
even if we investigate overlapping time windows of different
stocks because their returns are correlated as well. When daily
(close-to-close) price drops are examined we may get strong
cross-sample correlations if we include more than one event on one
particular day in our sample thus the t-values calculated by e.g.
\citeasnoun{atkins} may be biased. Our results show furthermore
that the t-values computed by \citeasnoun{park} and
\citeasnoun{bremer97} showing significant long-term  reversals (of
the length of 3-17 days) should be considered with care as well,
although significance of daily returns shown in these studies are
not affected by this problem, and are to be accepted.

Another result we obtain from the model data is that pre-event
returns are biased. Using the above sample selection process we
get returns significantly different from zero before the event.
This is due to the fact that during the last T minutes of the
pre-event period all events (i.e. large decreases) are excluded
but large increases are included: thus we have returns biased
toward price increases. The reason why this effect is so strong is
that the GARCH(1,1) process similarly to the real-market data
exhibits volatility clustering, i.e. large price decreases and
increases tend to cluster which only increases the number of
excluded pre-event large price decreases. It is most probable that
the sharp decline in price preceding intraday price jumps in our
results as shown in Table~\ref{ti} and on Figure~\ref{rev} are due
to this effect. This bias on the other hand should lead to a
similar price increase before intraday price drops, which however
cannot be observed; showing a strong asymmetry between the
pre-event market sentiment in case of large intraday increases and
decreases. Price drops seem thus to be preceded by strong selling.

%%%%%%%%%%%%%%%%%%%%%%%%%%%%%%%%%%%%%%%%%%%%%%%%%%%%%%%%%%%%%%%%
\vspace{1.5\textheight} % advance to the next page, but do not output figures

\small
%\begin{spacing}{1.8}
\bibliography{zak-abw}
%\end{spacing}

%%%%%%%%%%%%%%%%%%%%%%%%%%%%%%%%%%%%%%%%%%%%%%%%%%%%%%%%%%%%%%%%
\clearpage

\begingroup
  \parindent 0pt
  \parskip 2ex
  \def\enotesize{\normalsize}
  \theendnotes
\endgroup

\clearpage
\end{document}